\documentstyle[prl,eqsecnum,amsmath,amssymb,amsthm,aps,epsf,epsfig,psfig]{revtex}

\theoremstyle{remark}

\newcommand{\sg}[1]{\sigma_{#1}}
\newcommand{\om}[1]{\omega_{#1}}

\begin{document}
\bibliographystyle{/home/m/mx200/paper/nonlinear/fwm/jcp}
\draft
\title{
\large \bf
Hyperpolarizabilities for the one-dimensional infinite single-electron
periodic systems:\\
II. Dipole-dipole versus current-current correlations
}
\author{Minzhong Xu\thanks{Author to whom correspondence should be addressed. 
Email: mx200@nyu.edu}}
\address{
Department of Chemistry, New York University, New York, NY 10003
        }
\author{Shidong Jiang}
\address{
Department of Mathematical Sciences, New Jersey Institute of Technology,
Newark, NJ 07102
}

\date{\today}
\maketitle
\bigskip
\begin{abstract}
Based on Takayama-Lin-Liu-Maki model, analytical expressions for the
third-harmonic generation, DC Kerr effect, 
DC-induced second harmonic optical Kerr effect, optical Kerr effect or
intensity-dependent index of refraction and DC-electric-field-induced optical
rectification are derived under the static current-current($J_0J_0$) 
correlation for one-dimensional infinite chains. The results of 
hyperpolarizabilities under $J_0J_0$ correlation 
are then compared with those obtained using the dipole-dipole ($DD$)
correlation. The comparison shows that the conventional
$J_0J_0$ correlation, albeit 
quite successful for the linear case,
is incorrect for studying the nonlinear optical properties of periodic systems.\\
\end{abstract}



\section{introduction}
\label{sec:intro}
The different gauge approaches (${\bf p\cdot A}$ and ${\bf E\cdot r}$) have 
been 
adopted in the theoretical studies of both linear and nonlinear optical (NLO) 
properties for many materials\cite{shen,bloembergen,butcher}. For the 
current-current($JJ$) correlation (i.e., the ${\bf p\cdot A}$ gauge), most 
researchers tend to interpret the current operator $J$ as the static current
current $J_0$ \cite{bloembergen,butcher,wwu,mahan,maki}. For the linear 
transport theory, though the real part of $J_0J_0$ correlation causes the zero 
frequency divergence(ZFD), the convergent optical properties such 
as the linear susceptibility, the absorption coefficient, the linear
conductivity, etc could be obtained by using the imaginary part of $J_0J_0$ 
correlation alone, then by applying the Kramers-Kronig (KK) relations on the 
imaginary part of $J_0J_0$ correlation or including the diagmagnetic 
term\cite{mahan}. Hence, the static current-current($J_0J_0$) correlation 
is widely adopted in the linear transport 
theory\cite{mahan,maki,kivelson,gebhard,mxu3} and the ZFD is
often considered as a harmless technical nuisance and tacitly ignored by most
researchers. However, for the nonlinear case, $J_0J_0$ correlation\cite{mxu3}
encounters serious difficulties and the analytical results for nonlinear optical
properties do not even converge.

Among the polymer studies, theoretical calculations
of both linear\cite{maki,kivelson,gebhard,mxu3} and nonlinear optical 
properties\cite{wwu,agrawal,yuri,mxu1,mxu2,su1,su2,cwu1,cwu2,shuai,soos2}
have been carried out based on the different gauges
for the simplest $\pi$-conjugated polymers such as polyacetylene (PA). 
 For polyacetylene, some simple periodic, single electron, and
tight-binding approximation
models such as Su-Shrieffer-Heeger (SSH)\cite{ssh}
and Takayama-Lin-Liu-Maki (TLM)\cite{tlm} have
been established to interpret the experimental results\cite{heeger88}.
But in both linear and nonlinear calculations of the optical properties under 
the above models, as we pointed out recently\cite{mxu3}, there are 
some discrepancies between the conventional treatments using different gauges.
Specifically, if using the same set of wavefunctions but ignoring the
phase difference between both gauges and meanwhile applying the static current
in ${\bf p\cdot A}$ gauge, we cannot guarantee the equivalence between the two 
gauges, even though the final results look quite similar to each other 
qualitatively.
By the example calculation of linear susceptibility under SSH model for 
one-dimensional infinite chains, we strictly proved the nonequivalence 
between two gauges and ZFD could be
resolved by considering the gauge factor\cite{mxu3}. Since one needs to apply 
fairly complicated techniques to resolve ZFD in $J_0J_0$ correlation and 
preserve the equivalence between two gauges, we prefer the dipole-dipole($DD$)
correlation (i.e., the ${\bf E \cdot r}$ gauge) for nonlinear optical calculations 
for the polymers.

On one hand, the $DD$ correlation is derived by assuming a scalar
potential ${\bf E\cdot r}$ as perturbation, giving rise to the external
electric field ${\bf E}$. On the other hand, $J_0J_0$ correlation is
obtained by treating the time-dependent uniform vector potential ${\bf A}$ as 
perturbation. As long as one uses periodic boundary conditions, the scalar 
potential shows saw-shaped behavior and therefore the resulting electric field
is not uniform in the real space, while ${\bf J_0\cdot A}$ is uniform in real 
space. From this point of view, the $J_0J_0$ correlation 
seems  more appropriate than the $DD$ correlation. Thus it is our
interest to study some cases which avoid the ZFD difficulties in the
$J_0J_0$ correlation and reveal the pitfalls of the $J_0J_0$
correlation via a detailed  comparison between $DD$ and $J_0J_0$ correlations. 

Fortunately, the TLM model is one typical case that avoids the ZFD problem, 
although 
its sibling model - the SSH model is not\cite{yuri}. The static current operator 
$J_0$ derived from TLM model could give us the convergent results for 
hyperpolarizabilities when the frequency approaches 0. However, we
consider this result 
as a mere coincidence, since the linear 
susceptibility calculation based on TLM model diverges in the real part of 
$J_0J_0$ correlation\cite{mxu3}.
Nevertheless, we could use TLM model as a common ground to do the comparison 
between $DD$ and $J_0J_0$ correlations.

In \cite{jiang}, we have computed the analytical forms of 
hyperpolarizabilities for infinite chains by $DD$ correlation under both SSH 
and TLM models. In this paper, we first present a brief description of the
static current operator $J_0$ for both models and general
formulas for hyperpolarizabilities under $J_0J_0$ correlation in 
Section \ref{sec:theory}. We then proceed to carry out analytical 
calculations for DC Kerr effect, DC-induced second harmonic generation,
optical Kerr effect and DC-electric-field-induced optical rectification by 
$J_0J_0$ correlation under TLM model for infinite chains (Section \ref{sec:hyper}).
A detailed comparison of the results between $DD$ and 
$J_0J_0$ correlations is followed subsequently (Section \ref{sec:discu}).
The comparison shows that though there are some similarities for some
features such as resonant peaks and general shapes between these two
correlations, important and evident differences abound.
For instance, while $DD$ correlation clearly indicates the
nonexistence of the two-photon cusp in the third-harmonic 
generation(THG) spectrum \cite{mxu1,mxu2}, 
such cusp appeared in $J_0J_0$ correlation; and while $DD$ correlation 
obviously shows the break of overall permutation and Kleinman 
symmetries\cite{jiang}, $J_0J_0$ correlation maintains both the
overall permutation\cite{butcher} and Kleinman symmetries\cite{kleinman} 
for all frequencies. 
Finally, we present our conclusions in Section \ref{sec:concl}.

\section{Theory}
\label{sec:theory}
\subsection{nonlinear optical susceptibility under current-current correlation}
The $n$th-order nonlinear optical susceptibility under current-current($JJ$)
correlation is conventionally reduced to the static current-current($J_0J_0$)
correlation and defined as follows \cite{butcher,wwu,mxu3,mxu1}:
\begin{eqnarray}
&\chi&^{(n)}(\Omega; \omega_{1}, \ldots, \omega_{n})=-\delta_{n1}
\frac{n^{(e)}e^2}{\epsilon_0 m \omega_1^2} \hat{I}+
\displaystyle \frac{\chi_{j_0j_0}^{(n)}
( \Omega; \omega_{1}, \ldots, \omega_{n})}
{\epsilon_0 i\Omega \omega_1 \cdots \omega_n},
\label{eq:jj}
\end{eqnarray}
with $\displaystyle \Omega \equiv -\sum_{i=1}^{n} \omega_{i}$,
$n^{(e)}$ the electronic density, $m$ the electron electron mass,
$\epsilon_0$ the dielectric constant, $\hat{I}$
the unit dyadic, $\delta_{n,1}$ the Kronecker symbol, and
\begin{eqnarray}
&\chi&^{(n)}_{j_0j_0}(\Omega; \omega_{1}, \ldots, \omega_{n})= \frac{1}{n!V}
\left[ \frac{1}{\hbar} \right]^n \int  d{\bf r}_{1} \cdots d{\bf r}_{n}
\int dt_{1} \cdots dt_{n} \nonumber \\
& &\int d{\bf r} dt\, e^{-i {\bf k \cdot r}+ i \Omega t} \langle \hat{T}
{\bf \hat{J}_0} ({\bf r},t) {\bf \hat{J}_0}({\bf r}_{1},t_{1}) \cdots
{\bf \hat{J}_0} ({\bf r}_{n},t_{n}) \rangle,
\label{eq:j0j0}
\end{eqnarray}
where $V$ is the total volume, $\hat{T}$ is the time-ordering operator, and
${\bf \hat{J}_0}$ is the static current operator.

The Feynman diagram of $\chi^{(3)}$ is simply described as one connected
circle in the preceding paper (see Fig. 1 in \cite{jiang}).

\subsection{static current operator under SSH and TLM models}
The static current operator $\hat{J}_0$ could be found by the commutator 
between the dipole operator and Hamiltonian. For both SSH and TLM models, the 
current operators were derived in many previous 
works\cite{wwu,maki,kivelson,gebhard,mxu3}, here we only list the final results.

For the SSH model, under the same notation of the preceding 
paper\cite{jiang}, the static current operator $J_{0,SSH}$ is defined
by the formula
\begin{eqnarray}
\hat{J}_{0,SSH}=-\sum_{l,s} \displaystyle &i&\frac{e}{\hbar}
\left[ t_0+(-1)^l \frac{\Delta}{2} \right]
\left[ a-2(-1)^l u \right] \nonumber \\
& &(\hat{C}_{l+1,s}^{\dag}\hat{C}_{l,s}^{}-\hat{C}_{l,s}^{\dag}
\hat{C}_{l+1,s})^{},
\label{j0ssh}
\end{eqnarray}
where $t_0$ is the transfer integral between the nearest-neighbor sites,
$\Delta$ is the gap parameter and $\hat{C}_{l,s}^{\dag}(\hat{C}_{l,s})$
creates(annihilates) an $\pi$ electron at site $l$ with spin $s$.

For the TLM model (Eq.(2.2) in \cite{jiang}), by adopting the notation in 
Maki\cite{maki} and Wu\cite{wwu}'s work, the static current operator 
$J_{0,TLM}$ is defined by the formula
\begin{eqnarray}
\hat{J}_{0,TLM}=v_F\Psi^{\dag}(x)\sg{3}\Psi(x),
\label{j0tlm}
\end{eqnarray}
where $\Psi^{\dag}(x)=(\Psi^{\dag}_1(x),\Psi^{\dag}_2(x))$ is the
two-component spinor describing the left-going and right-going electrons,
$v_F$ is the Fermi velocity and $\vec{\sigma}$ are the Pauli matrixes.

As pointed out in our recent work\cite{mxu3}, detailed calculations show
that the above static current operators lead to the ZFD
in the linear response for both models. 
However, in the subsequent calculation for $\chi^{(3)}$, we show that the 
static current operator $J_0$ gives the convergent results for the TLM model. 
This provides us a convenient base to carry out the comparison of the
analytical results of $\chi^{(3)}$ 
between $DD$ and $J_0J_0$ correlations. Hence, the following calculations
are based on the TLM model only.

\section{Hyperpolarizabilities for TLM model under static current-current 
correlation}
\label{sec:hyper}
\subsection{General four-wave-mixing results}
We apply the general definition Eq.\eqref{eq:jj} and Eq.\eqref{eq:j0j0} to the
TLM model and obtain the following expression for 
$\chi^{(3)}_{TLM}(\Omega\equiv-(\om{1}+\om{2}+\om{3});\om{1},\om{2},\om{3})$
(or $\chi^{(3)}(\om{1},\om{2},\om{3})$ for short):
\begin{eqnarray}
\displaystyle
\chi^{(3)}_{TLM}(\Omega;\om{1},\om{2},\om{3})\equiv
\frac{\chi^{(3)}_{j_0j_0}}
{i\Omega\om{1}\om{2}\om{3}},
\label{eq:tlmx3}
\end{eqnarray}
where $\chi^{(3)}_{j_0j_0}$ is defined by the formula
\begin{eqnarray}
\chi^{(3)}_{j_0j_0}&=&\frac{-2e^4n_0v_F^4}{\hbar^3}\frac{1}{3!L}
\sum_{k,{\mathcal{P}}(\om{1},\om{2},\om{3})} \int
\frac{d\om{}}{2\pi}Tr\Biggl\{ \sg{3}G(k,\om{})
\sg{3}G(k,\om{}-\om{1}) \nonumber \\
&\sg{3}&G(k,\om{}-\om{1}-\om{2})
\sg{3}G(k,\om{}-\om{1}-\om{2}-\om{3})
\Biggr\}, \\
&\equiv&\frac{-2e^4n_0v_F^4}{\hbar^3L}\sum_k{\mathcal{S}}(\om{1},\om{2},\om{3})
\label{eq:tlmj0j0}
\end{eqnarray}
with L the chain length, ${\mathcal{S}}(\om{1},\om{2},\om{3})$ the 
summation of the permutations for $\om{1},\om{2},\om{3}$, and
Green's function $G$ defined by the formula
\begin{eqnarray}
G(k,s) = \frac{\omega-s+v_{F}k\sg{3}+\Delta\sg{1}/\hbar}
{(\omega-s)^{2}-\om{k}^{2}+i \epsilon}.
\label{eq:G}
\end{eqnarray}
In Eq. \eqref{eq:G}, $\omega_k$ is defined by the formula
\begin{eqnarray}
\om{k}=[(v_{F}k)^2+(\Delta/\hbar)^2]^{1/2}.
\end{eqnarray}

We now introduce the following three new variables:
\begin{equation}
c:=\Delta/\hbar,
\label{eq:c}
\end{equation}

\begin{equation}
x = \frac{\om{k}}{c}=\sqrt{1+(\frac{v_{F}\hbar}{\Delta}k)^2},
\end{equation}

\begin{equation}
z = \frac{\omega}{2c} = \frac{\hbar\omega}{2\Delta}.
\label{eq:z}
\end{equation}

Combining Eq. \eqref{eq:tlmj0j0}, Eq. \eqref{eq:c}-Eq. \eqref{eq:z} and
replacing the summation over $k$ by its continuous limit, we obtain
\begin{equation}\begin{aligned}\label{eq:chijj3}
\chi^{(3)}_{j_0j_0}(\om{1},\om{2},\om{3})&=\frac{e^4n_0v_{F}^{4}}{\pi\hbar^3}
\int_{-\infty}^{\infty}{\mathcal{S}}(\om{1},\om{2},\om{3})dk\\
&=\frac{2e^4n_0v_{F}^{3}}{\pi\hbar\Delta^2}
\int_{1}^{\infty}\frac{xdx}{\sqrt{x^2-1}}(c^3{\mathcal{S}}
(\om{1},\om{2},\om{3})).
\end{aligned}\end{equation}

Substituting Eq. \eqref{eq:chijj3} into Eq. \eqref{eq:tlmx3}, we have
\begin{equation}\begin{aligned}\label{eq:wuchi3}
\chi^{(3)}(\om{1},\om{2},\om{3})&=\frac{e^4n_0(\hbar v_{F})^{3}}{2^3\pi\Delta^6}
\frac{1}{-i(z_{1}+z_{2}+z_{3})z_{1}z_{2}z_{3}}
\int_{1}^{\infty}\frac{xdx}{\sqrt{x^2-1}}
(c^3{\mathcal{S}}(\om{1},\om{2},\om{3})),
\end{aligned}\end{equation}
where
\begin{equation}
z_{i} = \frac{\hbar \om{i}}{2\Delta}, \quad \text{$i=1\ldots 3$}.
\end{equation}

Eq.\eqref{eq:wuchi3} is the general formula for four-wave-mixing under 
$J_0J_0$ correlation. This is the same as defined in Wu's work\cite{wwu}.
As for nonlinear optical susceptibilities, there is no non-equilibrium
situation involved, the usage of Kelydish Green function in Wu's work is
not necessary.

Now Eq.\eqref{eq:wuchi3} is simplified to compute 
${\mathcal{S}}(\om{1},\om{2},\om{3})$ term. In this work, for the purpose of
comparing nonlinear response between different gauges,
we only obtain the analytical formats for third harmonic generation(THG), 
DC Kerr effect(DCKerr), DC-induced second harmonic generation(DCSHG), optical 
Kerr effect (i.e., intensity-dependent index of refraction (IDIR)), and 
DC-electric-field-induced optical rectification (EFIOR). The results under
$DD$ correlation with or without $\nabla_k$ contribution in the corresponding
figures are obtained from the preceding paper\cite{jiang}.

\subsection{Third Harmonic Generation 
$\chi^{(3)}(-3\om{};\om{},\om{},\om{})$}
\label{sec:hyper:thg}
Applying the Residue theorem and then using Maple to simplify 
${\mathcal{S}}(\om{},\om{},\om{})$ in the Eq.\eqref{eq:wuchi3}, we obtain:
\begin{equation}\begin{aligned}
{\mathcal{S}}(\omega,\omega,\omega)&=\frac{c^2}{\omega^4\om{k}}\left\{
\frac{4(5c^2-2\omega^2)}{3(4\om{k}^2-\omega^2)}
-\frac{8(c^2-\omega^2)}{3(\om{k}^2-\omega^2)}
+\frac{4(c^2-2\omega^2)}{(4\om{k}^2-9\omega^2)}\right\}\\
&=\frac{1}{2^4c^3z^4x}\left\{
\frac{(5-8z^2)}{3(x^2-z^2)}
-\frac{8(1-4z^2)}{3(x^2-4z^2)}
+\frac{(1-8z^2)}{(x^2-9z^2)}\right\},
\end{aligned}\label{eq:thg1}
\end{equation}

Combining Eq.\eqref{eq:wuchi3} and Eq. \eqref{eq:thg1}, we obtain
\begin{eqnarray}
\chi^{(3)}(\omega,\omega,\omega) =\frac{e^4n_0(\hbar
v_{F})^{3}}{1152\pi\Delta^6}\frac{1}{z^8}
\left\{3(1-8z^2)f(3z)-8(1-4z^2)f(2z)+(5-8z^2)f(z)\right\},
\label{eq:thg}
\end{eqnarray}
where the function $f(z)$ is defined by the formula
\begin{equation}\label{eq:fz}
f(z)=\int_{1}^{\infty}\frac{dx}{(x^2-z^2)\sqrt{x^2-1}}
\equiv \left \{
\begin{array}{lr}
\displaystyle  {\arcsin (z)\over z \sqrt{1-z^2}}  &(z^2<1),\\
\\
\displaystyle  -{\cosh^{-1} (z)\over z\sqrt{z^2-1}}+\displaystyle
{i\pi \over 2 z\sqrt{z^2-1}} &\ \ (z^2>1).
\end{array}
\right.
\end{equation}

As $z \rightarrow 0$, we have
\begin{equation}
\chi^{(3)}(\omega,\omega,\omega)=\frac{e^4n_0(\hbar
v_{F})^{3}}{\pi\Delta^6}
(\frac{4}{45}+\frac{32}{21}z^2+\frac{128}{7}z^4+\frac{18944}{99}z^6+O(z^8))
\end{equation}

Eq.\eqref{eq:thg} is exactly the same as Wu's result\cite{wwu}. 
By applying the following conversion between SSH and TLM model:
\begin{equation}
\label{eq:tlm_ssh}
\hbar v_F=2t_0a,
\end{equation}
then defining
\begin{equation}
\label{eq:x03}
\chi_{0}^{(3)}\equiv\frac{8}{45}\frac{e^4n_{0}(2t_0a)^3}{\pi\Delta^6},
\end{equation}
and choosing the same parameters as in the previous works\cite{mxu1,mxu2,jiang},
i.e. $\Delta=0.9 eV$, $n_0=3.2 \times 10^{14} cm^{-2}$ and $a=1.22 \AA$,
we obtain $\chi_0^{(3)} \approx 1.0 \times 10^{-10}$ esu.

\begin{figure}
\centerline{
\epsfxsize=9cm \epsfbox{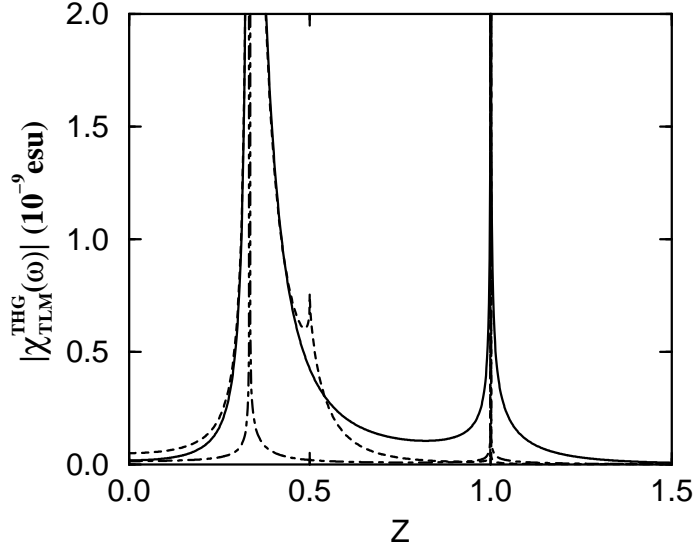}
}
\caption{The magnitude of third-harmonic generation(THG) under $J_0J_0$ 
correlation(dashed-line), under $DD$ correlation with(real line)
or without(dot-dashed) intraband contribution is in unit of
$10^{-9}$esu. $z\equiv\hbar\om{}/2\Delta$.}
\label{gr:thg}
\end{figure}

The magnitude of third-harmonic generation under $J_0J_0$ correlation, and 
that under $DD$ correlation with or without 
intraband contribution are plotted in Fig.\ref{gr:thg}. 
The theoretical discrepancies
of THG under different gauges have been noticed by many others' 
works\cite{mxu1,mxu2,su1,su2,cwu1,cwu2,shuai,soos2}. It has been addressed in
all works that the two-photon absorption peak observed in the experiments can 
not be explained by the single-electron models like SSH and TLM models.
Meanwhile, it was also pointed out that both gauges should reach the exact same
results if the calculations have been performed correctly. 
The reason why the discrepancy exists in the different gauges has not been
pinpointed in all previous calculations. Recently based on the same
models for the linear response\cite{mxu3}, we strictly proved that the gauge 
phase factor, which was ignored in the previous studies of optical 
properties, is the cause for the difference. When the gauge phase factor
is considered, the difference between different gauges could be 
resolved\cite{mxu3}.

\subsection{Optical Kerr effect or intensity-dependent index of refraction
$\chi^{(3)}(-\om{};\om{},-\om{},\om{})$}
Following an almost identical procedure of obtaining 
$\mathcal{S}(\om{},\om{},\om{})$, we have:
\begin{equation}\begin{aligned}
{\mathcal{S}}(\om{},-\om{},\om{})&=\frac{8c^2}{3}
\frac{(-48\om{k}^6+60\om{k}^4c^2+24\om{k}^4\om{}^2-35\om{k}^2c^2\om{}^2
-3\om{k}^2\om{}^4+2c^2\om{}^4)}{\om{k}^3(\om{k}^2-\om{}^2)(4\om{k}^2-\om{}^2)^3}\\
&=\frac{1}{6c^3}
\frac{(-12x^6+15x^4+24x^4z^2-35x^2z^2-12x^2z^4+8z^4)}
{x^3(x^2-4z^2)(x^2-z^2)^3}.
\end{aligned}\end{equation}

Following a similar procedure of evaluating $\chi^{(3)}(0,0,\omega)$, we
obtain the optical Kerr effect $\chi^{(3)}(-\om{};\om{},-\om{},\om{})$ 
(or $\chi^{(3)}(\om{},-\om{},\om{})$ for short) as follows:
\begin{equation}
\label{eq:idir}
\begin{aligned}
\chi^{(3)}(\om{},-\om{},\om{})=\frac{e^4n_0(\hbar v_{F})^{3}}{48\pi\Delta^6}
\frac{1}{z^8}
\left\{(4z^2-1)f(2z)-\frac{z^2(4z^2-1)}{2(1-z^2)^2}
-\frac{8z^6-12z^4+9z^2-2}{2(1-z^2)^2}f(z)\right\}.
\end{aligned}\end{equation}

As $z \rightarrow 0$, we have
\begin{equation}
\chi^{(3)}(\omega,-\omega,\omega)=\frac{e^4n_0(\hbar
v_{F})^{3}}{\pi\Delta^6}
(\frac{4}{45}+\frac{32}{63}z^2+\frac{128}{63}z^4+\frac{3584}{495}z^6+O(z^8))
\end{equation}

\begin{figure}
\centerline{
\epsfxsize=9cm \epsfbox{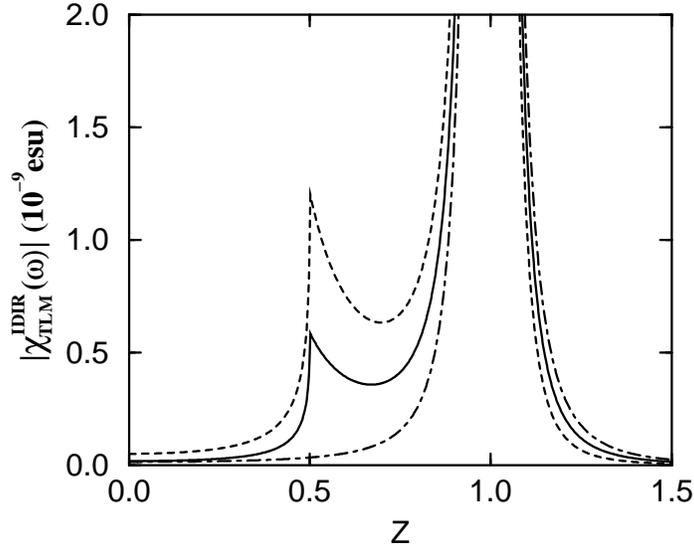}
}
\caption{ The magnitude of optical Kerr effect or 
intensity-dependent index of refraction(IDIR) under $J_0J_0$
correlation(dashed-line), under $DD$ correlation with(real line)
or without(dot-dashed) intraband contribution is in unit of
$10^{-9}$esu. $z\equiv\hbar\om{}/2\Delta$.}
\label{gr:idir}
\end{figure}

The magnitude of optical Kerr 
effect (i.e., intensity-dependent index of refraction(IDIR))
under $J_0J_0$ correlation, and that under $DD$ correlation with or without 
intraband contribution are plotted in Fig.\ref{gr:idir}. 

Eq.\eqref{eq:idir} is exactly the same as Eq.(13) in Wu's work\cite{wwu}. From  
Fig.\ref{gr:idir}, the results from $DD$ and $J_0J_0$ correlations all show the
cusp $z=1/2$. We would like to point out this is merely the van Hove 
singularity\cite{gebhard} by the
singular state density in one-dimensional polymer structure\cite{heeger88}, 
not the real resonant peak. Furthermore, the calculation through $DD$ 
correlations by
dropping $\nabla_k$ terms does not exhibit the $z=1/2$ cusp, showing
that the cusp is
related to the process of intraband-transition.

\subsection{DC Kerr effect $\chi^{(3)}(-\om{};0,0,\om{})$}
To evaluate the DC kerr effect $\chi^{(3)}(-\om{};0,0,\om{})$ (or 
$\chi^{(3)}(0,0,\omega)$ for short), 
we first evaluate 
$S(\omega_1,\omega_2,\omega_3)=S(z_1,z_2,z_3)$ for general
$z_i$ ($i=1,2,3$), then substitute it into Eq. \eqref{eq:wuchi3}
and take the limit $z_1\rightarrow 0$, $z_2\rightarrow 0$, and
$z_3\rightarrow z$. We obtain
\begin{equation}
\chi^{(3)}(0,0,\om{})=\frac{e^4n_0(\hbar v_{F})^{3}}{576\pi\Delta^6}
\frac{1}{z^4(z^2-1)^3}
\left\{3(3-8z^2)f(z)+(16z^6-40z^4+18z^2-9)\right\}
\label{eq:dckerr}
\end{equation}
As $z \rightarrow 0$, we have
\begin{equation}
\chi^{(3)}(0,0,\om{})=\frac{e^4n_0(\hbar v_{F})^{3}}{\pi\Delta^6}
(\frac{4}{45}+\frac{16}{63}z^2+\frac{32}{63}z^4+\frac{256}{297}z^6+O(z^8))
\end{equation}

\begin{figure}
\centerline{
\epsfxsize=9cm \epsfbox{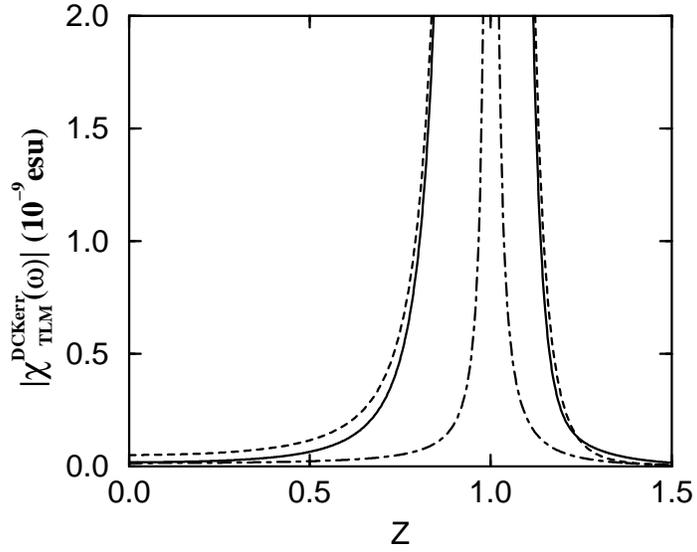}
}
\caption{ The magnitude of DC Kerr effect(DCKerr) under $J_0J_0$
correlation(dashed-line), under $DD$ correlation with(real line)
or without(dot-dashed) intraband contribution is in unit of
$10^{-9}$esu. $z\equiv\hbar\om{}/2\Delta$.}
\label{gr:dckerr}
\end{figure}

The magnitude of DC Kerr 
effect(DCKerr) under $J_0J_0$ correlation, and that 
under $DD$ correlation with or without 
intraband contribution are plotted in Fig.\ref{gr:dckerr}. The figure only 
shows one resonant peak at $z=1$ for all 3 cases.

\subsection{DC induced second harmonic generation 
$\chi^{(3)}(-2\om{};0,\om{},\om{})$}
Following a similar procedure of evaluating $\chi^{(3)}(0,0,\omega)$, we
obtain the DC induced second harmonic generation
$\chi^{(3)}(-2\om{};0,\om{},\om{})$ (or $\chi^{(3)}(0,\om{},\om{})$
for short) as follows:
 
\begin{equation}
\chi^{(3)}(0,\om{},\om{}) =\frac{e^4n_0(\hbar
v_{F})^{3}}{384\pi\Delta^6}
\frac{1}{z^8}\left\{
\frac{2-9z^2}{4z^2-1}f(2z)
+\frac{z^2(1+z^2-8z^4)}{(z^2-1)(4z^2-1)}
-\frac{2}{z^2-1}f(z)\right\}
\label{eq:dcshg}
\end{equation}

As $z \rightarrow 0$, we have
\begin{equation}
\chi^{(3)}(0,\omega,\omega)=\frac{e^4n_0(\hbar
v_{F})^{3}}{\pi\Delta^6}
(\frac{4}{45}+\frac{16}{21}z^2+\frac{32}{7}z^4+\frac{11776}{495}z^6+O(z^8))
\end{equation}

\begin{figure}
\centerline{
\epsfxsize=9cm \epsfbox{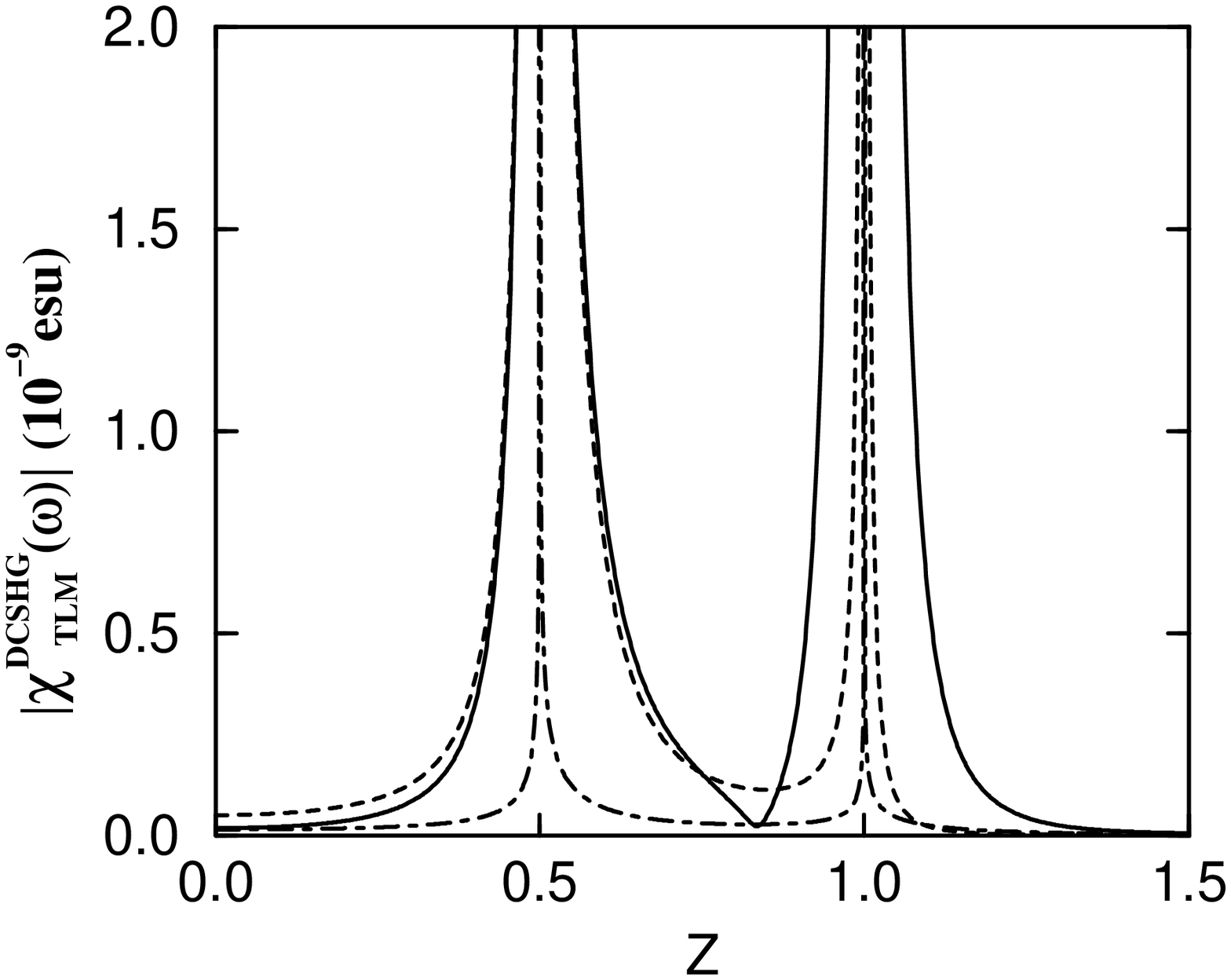}
}
\caption{ The magnitude of DC induced second harmonic generation(DCSHG)
under $J_0J_0$
correlation(dashed-line), under $DD$ correlation with(real line)
or without(dot-dashed) intraband contribution is in unit of
$10^{-9}$esu. $z\equiv\hbar\om{}/2\Delta$.}
\label{gr:dcshg}
\end{figure}

The magnitude of DC induced second 
harmonic generation(DCSHG)
under $J_0J_0$ correlation, and that under $DD$ correlation with or without 
intraband contribution are plotted in Fig.\ref{gr:dcshg}. The Figure clearly
shows two resonant peaks at $z=1/2$ and $z=1$. The width of $z=1$ peak 
suggests that the peak will not be so huge under $J_0J_0$ correlation than 
$DD$ correlation if the damping is included.

\subsection{DC-electric-field-induced optical rectification
$\chi^{(3)}(0;\om{},-\om{},0)$}
\begin{figure}
\centerline{
\epsfxsize=9cm \epsfbox{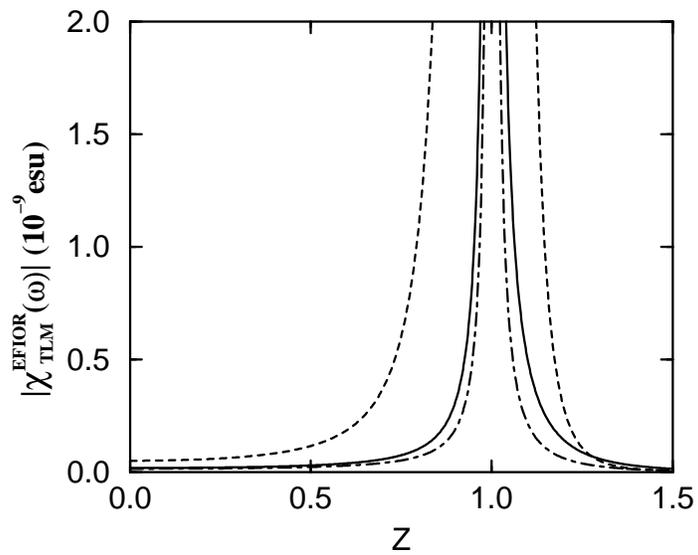}
}
\caption{ The magnitude of DC-electric-field-induced optical 
rectification(EFIOR) under $J_0J_0$ correlation(dashed-line), 
under $DD$ correlation with(real line)
or without(dot-dashed) intraband contribution is in unit of
$10^{-9}$esu. $z\equiv\hbar\om{}/2\Delta$.}
\label{gr:efior}
\end{figure}

After the calculations, we obtain the same results as those in DC Kerr effect.
Kleinman symmetry\cite{kleinman} is preserved in this calculation for all 
regions. This result is different from $DD$ correlation since $J_0J_0$ 
correlation maintains the commuting feature for all operators.
Due to the nonequivalence between EFIOR and DCKerr under $DD$ correlation, we still 
plot the magnitude of EFIOR under $J_0J_0$ correlation, and that 
under $DD$ correlation with or without intraband contribution in 
Fig.\ref{gr:efior}.

\section{discussions}
\label{sec:discu}
\subsection{Nonequivalence between $DD$ and $J_0J_0$ correlations} 
From the above calculations in Sec. \ref{sec:hyper}, the nonequivalence of
hyperpolarizabilities between $DD$ and $J_0J_0$ correlations can be found in
all results, though there are some similarities in the resonant peak,
the shape of the curve, etc. To understand the difference between the gauges
in the models, we present a possible explanation in our previous 
work\cite{mxu3}. To maintain the self-completeness of this work, we 
also briefly address the explanation here.

If the electromagnetic field is applied, the Sch\"{o}dinger equation is given by:
\begin{eqnarray}
i\hbar\displaystyle\frac{\partial}{\partial t}\psi({\bf r},t)=
\left[\displaystyle\frac{1}{2m}(\hat{\bf p}-q {\bf A})^2+V({\bf r})
+q\phi\right]\psi({\bf r},t),
\label{eq:se}
\end{eqnarray}

where $\psi({\bf r},t)$ is the exact wave function at space position ${\bf r}$
and time $t$, $m$ is the particle mass, $q$ is the
electrical charge, $V({\bf r})$ is the potential, and ${\bf A}$ and $\phi$ are
vector and scalor potential,
respectively. Suppose now  ${\bf A}$ and $\phi$ undergo the following transformation:\\
\begin{equation}
\left \{
\begin{array}{l}
{\bf A} \to {\bf A'}={\bf A}+\nabla f({\bf r},t)\\
\phi \to \phi'=\phi-\displaystyle \frac{\partial}{\partial t}
f({\bf r},t),
\end{array}
\right.
\label{eq:gt1}
\end{equation}
where $f({\bf r},t)$ is arbitrary, and ${\bf A'}$ and $\phi'$ are new vector
and new scalor potentials after the transformation
Eq.(\ref{eq:gt1}). Then it can
be shown\cite{cohen} that the form of the Sch\"{o}dinger equation will be
exactly the same if the old wave function $\psi$ makes the following change
into the new exact wave function $\psi'$:
\begin{eqnarray}
\psi \to \psi'=e^{iF_g({\bf r},t)}\psi=\hat{T}_G({\bf r},t)\psi,
\label{eq:gt2}
\end{eqnarray}
where gauge phase factor $F_g({\bf r},t)$ is defined as:
\begin{eqnarray}
F_g({\bf r},t) \equiv \displaystyle \frac{q}{\hbar}f({\bf r},t).
\label{eq:gf}
\end{eqnarray}

The above Eq.(\ref{eq:gt1}) and Eq.(\ref{eq:gt2}) are called gauge
transformation (or $U(1)$ transformation\cite{fradkin}).

By utilizing the long-wavelength approximation\cite{butcher,mahan}, the 
electric field ${\bf E}$ is described as ${\bf E}={\bf E_0} e^{-i\omega t}$.

If we consider the following initial scalor and vector potentials
under ${\bf E \cdot r}$ gauge:
\begin{eqnarray}
{\bf A}=0, \phi=-{\bf E} \cdot r.
\label{eq:er}
\end{eqnarray}
After choosing the gauge phase factor $F_g$ as
\begin{eqnarray}
F_g=\displaystyle \frac{q{\bf E \cdot r}}{i\hbar\omega}=\displaystyle
\frac{q}{\hbar}{\bf A'\cdot r},
\label{eq:fg}
\end{eqnarray}
by Eq.(\ref{eq:gt1}), we obtain the new vector and new scalor potential under
${\bf p \cdot A}$ gauge as:
\begin{eqnarray}
{\bf A'}=\displaystyle\frac{\bf E}{i\omega}, \phi'=0.
\label{eq:pa}
\end{eqnarray}
The connection between the old and new wave function is determined by
Eq.(\ref{eq:gt2}).

Under perturbative schemes to study the optical response,
conventionally people use the exact same set of unperturbed wave functions
$\psi^0_n({\bf r},t)$ of Hamiltonian $\hat{H}_0$ (when ${\bf A}=0$ and
${\phi=0}$ in Eq.(\ref{eq:se})) to serve as our expansion basis for both
${\bf E \cdot r}$ and ${\bf p \cdot A}$
gauges\cite{bloembergen,butcher}. However, we should
point out that the wave functions for both ${\bf E \cdot r}$ and
${\bf p \cdot A}$ gauges (before and after gauge transformation)
should also be restricted by the gauge phase factor $F_g$ from
Eq.(\ref{eq:gt2}), therefore two basis sets for both gauges are {\bf not}
the exact same unperturbated wave functions $\psi^0_n({\bf r},t)$, but
are different by the gauge phase factor $F_g$. And the Hamiltonian under
two gauges (${\bf E\cdot r}$ and ${\bf p \cdot A}$) are not necessary
equivalent if they are treated independently and are isolated from the
connection between the wave functions under the two gauges. Unfortunately,
this crucial point has not been clearly illustrated and obviously missed
by previous works using perturbation schemes \cite{shen,bloembergen,butcher}. Especially
under current-current correlation scheme, the gauge phase factor's
contribution is obviously ignored
and $A^2(t)$ term is considered of no physical meanings\cite{butcher}.
Thus the current-current correlation is conventionally reduced into the
$J_0J_0$ formula such as Eq.(\ref{eq:jj}), and the equivalence
between current-current and dipole-dipole correlations is usually considered
as $J_0J_0$ and $DD$ correlations under the exact same basis of
unperturbed wave functions\cite{bloembergen,butcher,wwu}.

Langhoff, Epstein and Karplus covered the topics of time-dependent 
perturbative theory\cite{langhoff} and sharply pointed
out that the time-dependent phase in wave functions is very essential, the
improper treatment of time-dependent phase will cause secular divergence in
time-dependent perturbations. In field theory, it is also well-understood that
the improper treatment of the phase factor will cause divergence\cite{mahan}.
Since the gauge phase factor Eq.(\ref{eq:fg}) is obviously time-dependent,
neglecting this phase factor will cause the ZFD in the susceptibility
computations.

Generally speaking, the widely-adopted conventional formula 
under $J_0J_0$ is incorrect. It ignores both the gauge phase factor's 
influence and diamagnetic term's contribution\cite{mxu3}. For the linear case,
we strictly proved that after taking the consideration of the diamagnetic term 
and the gauge phase factor, both $DD$ and $J_0J_0$ correlations yield the exact
the same result for both SSH and TLM models. The details could be found 
in \cite{mxu3} and will not 
be repeated here. But for the nonlinear case as we mentioned in the 
THG calculations\cite{mxu1,mxu2}, the complexity to include the 
gauge phase factor in $JJ$ correlation suggested $DD$ correlation may be more
suitable for further studies.

\subsection{Zero frequency divergence (ZFD)}
In general, $J_0J_0$ correlation leads to the zero frequency divergence
in the nonlinear optical studies. 
The static current operator in the TLM model
coincidentally avoids ZFD problem in the nonlinear calculations shown above, 
which does not mean that it is flawless. For example, linear 
calculation based on $J_0$ in TLM model leads a ZFD problem\cite{mxu3}. 
By splitting the $J_0$ term into inter- and intra-band currents in the TLM model
and performing the nonlinear calculations to determine the contributions
from two different currents, we find that the hyperpolarizabilities for
both cases have ZFD. For the SSH model, the static current operator 
$J_0$\cite{gebhard} leads to the ZFD in
nonlinear calculations\cite{yuri}. If the gauge phase factor\cite{mxu3} is 
properly considered in our calculations, the ZFD problem could be resolved.
Therefore, the ZFD problem for nonlinear calculations under the conventional 
schema of Eq.\eqref{eq:jj} and Eq.\eqref{eq:j0j0} is not just a
technical nuisance which can be tacitly ignored.

\subsection{The overall permutation and Kleinman symmetries}
Based on the ${\bf p\cdot A}$ gauge, the general formulas of 
$J_0J_0$ correlations\cite{butcher,wwu} preserve both the overall
permutation\cite{butcher} and Kleinman symmetry\cite{kleinman} of 
hyperpolarizabilities in any systems. Without surprise,
our calculations of hyperpolarizabilities under $J_0J_0$ correlation preserve 
both the overall permutation and Kleinman symmetries. However, the
overwhelming majority of recent experiments on various physical systems generally 
refute Kleinman symmetry\cite{dailey}. Based on ${\bf E\cdot r}$ gauge and
1D periodic models, we analytically showed the break of overall permutation 
and Kleinman symmetry\cite{jiang,mxu4}. Therefore, the experimental testing on
the overall permutation symmetry in periodic systems can also be used as
a valid test for the conventional $J_0J_0$ correlation and ${\bf p\cdot A}$ 
gauge. Detailed discussions of the symmetry break and some suggested 
experiments could be found in \cite{jiang,mxu4}.

\section{Conclusions}
\label{sec:concl}
For the infinite chains under TLM model, the analytical solutions of THG, 
DCKerr, DCSHG, IDIR and 
EFIOR are obtained through $J_0J_0$ correlation. The results are not equivalent
to those under $DD$ correlations\cite{jiang}. It shows that the conventional 
$J_0J_0$ 
correlation formula is incorrect for studying the nonlinear optical properties.
Considering the complexity of including the gauge phase factor and other terms
for the current-current correlation, $DD$ correlation may be much more suitable
in the nonlinear studies. 

\bibliography{}

\begin{thebibliography}{10}
\bibitem{shen}Y.R. Shen, {\it The Principles of Nonlinear Optics} (John Wiley
$\&$ Sons, Inc, 1984).
\bibitem{bloembergen}N. Bloembergen, {\it Nonlinear Optics} (W.A. Benjamin,
Inc, 1977).
\bibitem{butcher}P.N. Butcher and D. Cotter, {\it The Elements of Nonlinear
Optics} (Cambridge University Press, 1990).
\bibitem{wwu}Weikang Wu, Phys. Rev. Lett. {\bf 61}, 1119 (1988).
\bibitem{mahan} G D Mahan, {\it Many-Particle Physics} (New York: Plenum).
\bibitem{maki} K Maki and M Nakahawa, Phys. Rev. B {\bf 23}, 5005 (1981).
\bibitem{kivelson} S. Kivelson, T. K. Lee, Y R Lin-Liu, I Peschel and L Yu,
Phys. Rev. B {\bf 25} 4173 (1982).
\bibitem{gebhard} F. Gebhard, K Bott, M Scheidler, P Thomas and S W Koch, 
Phil. Mag. B {\bf 75} 1 (1997).
\bibitem{mxu3} M.Z. Xu and X. Sun, J. Phys. Condens. Matter {\bf 11}, 
9823 (1999).
\bibitem{agrawal} G. P. Agrawal, C. Cojan, and C. Flytzainis, Phys. Rev. B
{\bf 17}, 776 (1978).
\bibitem{yuri}Y.I. Dakhnovskii and K.A. Pronin, Synth. Met. {\bf 54}, 295
(1993).
\bibitem{mxu1} M.Z. Xu and X. Sun, Phys. Rev. B {\bf 61}, 15766 (2000).
\bibitem{mxu2} M.Z. Xu and X. Sun, Phys. Lett. A {\bf 257}, 215 (1999);
{\bf 259}, 502 (1999).
\bibitem{su1}J. Yu, B. Friedman, P.R. Baldwin, and W.P. Su, Phys. Rev. B
{\bf 39}, 12814 (1989).
\bibitem{su2} J. Yu and W.P. Su, Phys. Rev. B {\bf 44}, 13315 (1991).
\bibitem{cwu1}C.Q. Wu and X. Sun, Phys. Rev. B {\bf 41}, 12845 (1990).
\bibitem{cwu2}C.Q. Wu and X. Sun, Phys. Rev. B {\bf 42}, R9736 (1990).
\bibitem{shuai}Z. Shuai and J.L. Br\'{e}das, Phys. Rev. B {\bf 44}, R5962
(1991).
\bibitem{soos2}F.C. Spano and Z.G. Soos, J. Chem. Phys. {\bf 99}, 9265 (1993).
\bibitem{ssh} W.P. Su, J.R. Schrieffer, and A.J. Heeger, Phys. Rev. Lett.
{\bf 42}, 1698 (1979); Phys. Rev. B {\bf 22}, 2099 (1980).
\bibitem{tlm} H. Takayama, Y.R. Lin-Liu, and K. Maki, Phys. Rev. B {\bf 21},
2388 (1980).
\bibitem{heeger88}A.J. Heeger, S. Kivelson, J.R. Schrieffer, and W.P. Su, Rev.
Mod. Phys. {\bf 60}, 781 (1988) and references there in.
\bibitem{jiang} S.D. Jiang and M.Z. Xu, the proceeding paper.
\bibitem{kleinman}D.A. Kleinman, Phys. Rev. {\bf 126}, 1977(1962).
\bibitem{cohen} C. Cohen-Tannoudji, J. Dupont-Roc, and G. Grynberg,
{\it Photons and Atoms} (John Wiley \& Sons, Inc, 1989).
\bibitem{fradkin}E. Fradkin, {\it Field Theories of Condensed Matter Systems}
(Addison-Wesley Publishing Company, 1991), p.9.
\bibitem{langhoff} P.W. Langhoff, S.T. Epstein, and M. Karplus, Rev. Mod.
Phys. {\bf 44}, 602 (1972).
\bibitem{dailey} C.A. Dailey, B.J. Burke and G.J. Simpson, Chem. Phys.
Lett. {\bf 390}, 8 (2004) and references there in.
\bibitem{mxu4} M.Z. Xu and S.D. Jiang, cond-mat/0505307.
\end{thebibliography}

\end{document}